\documentclass[conference]{IEEEtran}
\IEEEoverridecommandlockouts

\makeatletter
\def\markboth#1#2{\def\leftmark{\@IEEEcompsoconly{\sffamily}\MakeUppercase{\protect#1}}%
\def\rightmark{\@IEEEcompsoconly{\sffamily}\MakeUppercase{\protect#2}}}
\makeatother

\usepackage[inline,ignoremode]{trackchanges}

\tcignore{\ref}{1}{1}
\tcignore{\eqref}{1}{1}
\tcignore{\cite}{1}{1}

\addeditor{Felipe}
\addeditor{Sundeep}
\addeditor{Elza}

\usepackage[utf8x]{inputenc}
\usepackage[english]{babel}
\usepackage{ucs}
\usepackage{amsmath}
\usepackage{amsfonts}
\usepackage{amssymb}
\usepackage{amsthm}
\usepackage{array}
\usepackage{verbatim}
\usepackage{listings}
\usepackage{stfloats}

\usepackage{algorithm}
\usepackage{algorithmic}
\usepackage{url}
  \usepackage{enumerate}
\usepackage{cite}
\usepackage{etoolbox}

\ifCLASSINFOpdf
  \usepackage[pdftex]{graphicx}
  \usepackage{subfigure}
  \usepackage{epstopdf}
   \graphicspath{{./img/}}
   \DeclareGraphicsExtensions{.eps,.ps}
\else
  \usepackage[dvipdf]{graphicx}
  \graphicspath{{./img/}}
   \DeclareGraphicsExtensions{.eps,.ps}
\fi

\newtoggle{conference}
\toggletrue{conference} 

\newcommand{\x}{\mathbf{x}}

\theoremstyle{definition}
\newtheorem{definition}{Definition}
\theoremstyle{plain}
\newtheorem{lemma}{Lemma}
\newtheorem{theorem}{Theorem}

\theoremstyle{remark}
\newtheorem{remark}{Remark}

\title{Scaling Laws for Infrastructure Single and
Multihop Wireless Networks in Wideband Regimes}

\author{
  \IEEEauthorblockN{Felipe G\'omez-Cuba}
    \IEEEauthorblockA{AtlantTIC, University of Vigo\\
      C.P. 36310 Vigo, España\\
      {\tt fgomez@gti.uvigo.es } }
   \and
  \IEEEauthorblockN{Sundeep Rangan and Elza Erkip}
    \IEEEauthorblockA{NYU Polytechnic School of Engineering, \\
      Brooklyn, NY 11201, USA\\
      {\tt \{srangan,elza\}@poly.edu}
      }
\thanks{This reasearch was partially supported by FPU12/01319 MECD, University of Vigo and Xunta de Galicia, Spain; NSF-1237821 and NSF-1302336.
}
}

\begin{document}
\maketitle
\begin{abstract}
With millimeter wave bands emerging as a strong candidate for 5G cellular networks, next-generation systems may be in a unique position where spectrum is plentiful.
To assess the potential value of this spectrum, this paper derives scaling laws on the per mobile downlink feasible rate with large bandwidth and number of nodes, for both Infrastructure Single Hop (ISH) and Infrastructure Multi-Hop (IMH) architectures.
It is shown that, for both cases, there exist \emph{critical bandwidth scalings} above which increasing the bandwidth no longer increases the feasible rate per node.
These critical thresholds coincide exactly with the bandwidths where, for each architecture, the network transitions from being degrees-of-freedom-limited to power-limited.
For ISH, this critical bandwidth threshold is lower than IMH when the number of users per base station grows with network size.
This result suggests that multi-hop transmissions may be necessary to fully exploit large bandwidth degrees of freedom in deployments with
growing number of users per cell.
\end{abstract}


\section{Introduction}

\label{sec:introduction}
New technologies in the millimeter wave (mmW) bands between 30 and 300~GHz may move networks to a new regime where spectrum is up to 200 times current cellular allocations, with further spatial degrees of freedom made available by very high-dimensional antenna arrays \cite{Rappaport2013,RanRapEr:14}.
From a theoretical perspective, the possibility of vast quantities of spectrum raises two broad fundamental questions: What is the ultimate value of
very large bandwidths, and how can networks be best designed to exploit them?

It is well-known that links in wideband regimes become limited by power rather than bandwidth, particularly with channel fading \cite{journals/tit/MedardG02,journals/twc/LozanoP12}.
In these scenarios, \emph{multi-hop relaying} transmissions can increase the received power by shortening the distance of each link, thereby potentially increasing the overall capacity.
Unfortunately, in current cellular network implementations, the gains of multi-hop relaying have been limited \cite{peters2009relay}: Since links tend to be bandwidth-limited, the benefits of increased received power from relaying are generally outweighed by the loss in degrees of freedom from half-duplex constraints. However, in wideband regimes where power is the dominant constraint and degrees of freedom are plentiful, multi-hop relaying may provide much greater value.

To assess this hypothesis, this paper evaluates the performance of multi-hop relaying through the framework of scaling laws, following the classic analysis of Gupta and Kumar~\cite{kumar2003capacity}.
Specifically, we consider a sequence of wireless networks undergoing fading, indexed by the number of single antenna nodes, $n$.
The number of base stations (BSs), BS antennas, network deployment area, and bandwidth are assumed to scale by a function of $n$. The goal is to evaluate how the per node data rate $R(n)$ that is simultaneously achievable at all nodes, which we call \textit{feasible rate}, scales. For ease of exposure, we only consider downlink transmissions; similar results for uplink can be derived and will be reported elsewhere.
Within this framework, we compare two protocols:
Infrastructure single hop (ISH), where transmissions are made directly between the BSs and nodes as in traditional cellular networks
today, and infrastructure multi-hop (IMH) where communication takes place through a sequence of relayed transmission with user nodes acting as relays. Analysis of upper bounds, hierarchical cooperation protocols, and comparison with ad-hoc networks are left for future work.

We derive new scaling laws that govern the feasible rate $R(n)$, for both ISH and IMH in the downlink. The scaling laws reveal that there are three key regimes of operation:
\begin{enumerate}
\item When bandwidth per node scaling is relatively small, both ISH and IMH are degrees-of-freedom-limited and their rates scale with bandwidth.

\item When bandwidth scaling per node is sufficient to make ISH transmissions power limited, ISH cannot exploit the increasing bandwidth. However, the
    IMH protocol may still exploit the full bandwidth when the number of nodes scales faster than number of BS.
\item Finally, for very large bandwidth scalings, both IMH and ISH become power-limited and $R(n)$ ceases to increase with bandwidth for either system. However, the exponent of $R(n)$ as a function of $n$ for IMH remains above that for ISH when the number of nodes scales faster than number of BS.
\end{enumerate}	
The existence of these three regimes has immediate consequences on the capacity of next-generation networks.
Traditional cellular systems with very limited bandwidth will remain in the first regime -- confirming the general observations that relaying provides few benefits in current bandwidth-constrained networks. On the other hand, mmW systems with very wide bandwidths and large number of antennas per base station can potentially enter the second or third regimes.
Thus, multi-hop relaying may play a valuable role in cellular network evolution with increasing number of nodes per BS if very large bandwidths are used, as in mmW.

\subsection{Prior Work}

\iftoggle{conference}{
The original Gupta-Kumar result \cite{kumar2003capacity} considered the capacity
of \emph{ad-hoc} networks with no fixed infrastructure.  Subsequent
works considered hierarchical cooperation \cite{ozgur2009information} along with
mobility, broadcast, and infrastructure; see \cite{Lu2013} for a comprehensive review. Our model for networks with infrastructure is based on \cite{journals/tit/ShinJDVCLT11}. We model infrastructure similarly, but we only consider a simpler node data flow model to and from the BS separately.
}{

The original Gupta-Kumar result \cite{kumar2003capacity} showed that the feasible rate in an \textit{ad-hoc} network scales as $R(n)\propto\Theta(\frac{1}{\sqrt{n}})$\footnote{The notation $f(n)\propto\Theta(g(n))$ means that $\exists$ two positive real constants $c,c'\in\mathbb{R}^+$ and an index $n_0\in\mathbb{N}$ such that $f(n)$ is lower and upper bounded by a scaled version of $g(n)$, $cg(n)<f(n)\leq c'g(n)$ $\forall n>n_0$}
where $n$ is the number of nodes.  Hence, the per node capacity decreases with the network size.  Subsequent works improved this result. For instance, hierarchical cooperation \cite{ozgur2009information} may achieve linear scaling (i.e. $R(n) = \Theta(1)$) in ad-hoc networks. Some results, like \cite{ozgur2009information}, are valid for high, yet finite, values of $n$, but cease to hold when $n$ is so high that the finite number of degrees of freedom in the electromagnetic field becomes the ultimate physical limitation, forcing $R(n)\leq\Theta(\frac{\log(n)^2}{\sqrt{n}})$ \cite{Franceschetti2009}. As in \cite{ozgur2009information}, our results are valid in the absence of this physical limitation.

Following the classic result of Gupta and Kumar, there have been extensions of scaling laws in ad-hoc networks introducing cooperation, mobility, broadcast, infrastructure, see \cite{Lu2013} for a comprehensive review. Our model for networks with infrastructure is based in \cite{journals/tit/ShinJDVCLT11}. We model infrastructure similarly, but we only consider a simpler node data flow model to and from the BS separately.
}

The main novelty of our model is including the impact of very large bandwidths in infrastructure capacity scaling. Most scaling analyses consider a finite bandwidth; however in such a setup overspreading as in \cite{journals/tit/MedardG02} never occurs, because bandwidth may be sliced arbitrarily thin as the number of nodes grows. Another approach would be to let $W\rightarrow\infty$ for every finite $n$, and \textit{then} let $n$ grow, as in \cite{Negi2004}; but this forces the network to be always power-limited and provides no insights in the interplay between critical bandwidth and network architecture. To find out what happens in between, we let $W$ and $n$ increase with $n$ polynomially with an arbitrary exponent
\begin{equation}
    \psi:=\lim_{n,W\rightarrow \infty}\frac{\log W}{\log n},
\end{equation}
where the two special cases above are $\psi=0$ and $\psi=\infty$. By introducing this new parameter, bandwidth scaling becomes $W=W_0n^{\psi}$, and different protocols can be compared in terms of how the critical bandwidth scales.

A key feature of our analysis is that we can model the power overhead for
channel estimation in the wideband regime.
Recent experimental measurements have demonstrated that mmW outdoor links
often rely on diffuse reflections with multiple NLOS paths \cite{Rappaport2013,AkLiuRanRapEr:13-arxiv}.
These diffuse reflections will present new channel coefficients
as bandwidth grows, so the wideband fading models described in \cite{journals/tit/MedardG02,journals/twc/LozanoP12} apply. M{\'e}dard and Gallager \cite{journals/tit/MedardG02} showed that mutual information decreases to zero as channel estimation errors increase when a finite power is spread over an excessive bandwidth (i.e.\ \textit{overspreading}). Lozano and Porrat \cite{journals/twc/LozanoP12} argued that for non-peaky\footnote{Signals with finite fourth moment, or equivalently a finite-power signal that is not constituted by averaging over time a flash pulse of infinite instantaneous power.} signaling the maximum achievable rate lies within a constant factor to the capacity of the AWGN channel at infinite-bandwidth, and the \textit{critical bandwidth} where this happens grows linearly in power.
Our feasible rate estimates account for these
critical bandwidth and overspreading effects.

\section{System Model}
\label{sec:model}
\begin{figure}[!t]
 \centering
 \includegraphics[width=0.8\columnwidth]{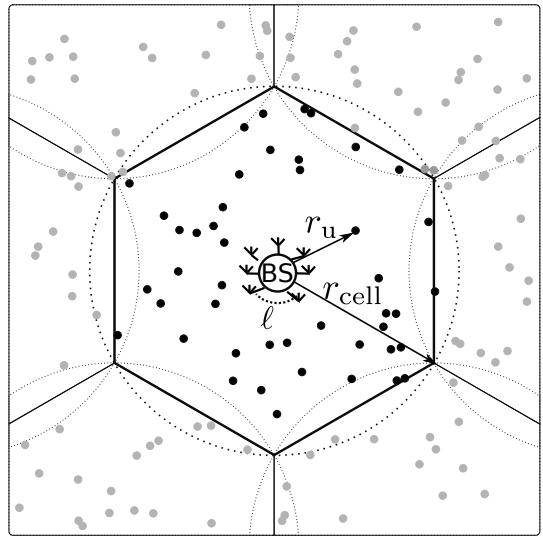}
 \caption{Detail of network model, only one cell and its neighbors are shown.}
 \label{fig:model}
\end{figure}

We consider an infrastructure cellular wireless network as in Fig. \ref{fig:model}, with $n$ single-antenna nodes uniformly distributed in area $A$. The network is supported by $m$ infrastructure nodes, or BSs, with $\ell$ antennas each, and communication takes place over an increasing bandwidth $W$. Nodes have transmit power constraint $P$ and BSs have power constraint $P_{\mathrm{BS}}$. The network is organized regularly in cells with radius $r_{\mathrm{cell}}$. Signal attenuation with distance follows path-loss exponent $\alpha$ and channels experience random small scale fading (not necessarily Rayleigh, as overspreading occurs for any distribution \cite{journals/tit/MedardG02}). Channel state information is not a priori available to the terminals. We only study the downlink from the BS to the nodes; uplink can be considered similarly.

We study the feasible rate $R(n)$ scaling as $n\rightarrow\infty$. The scaling relation between $n$ and network parameters is defined in Table \ref{tab:exponents}. The exponents of the number of BSs and BS antennas follow from \cite{journals/tit/ShinJDVCLT11}. The constraint $\beta+\gamma<1$ ensures that the number of infrastructure antennas per node does not grow without bounds. The scaling of the network area is as proposed by \cite{ozgur2009information} to model a continuum between \textit{dense} ($\nu=0$) and \textit{extended} ($\nu=1$) networks. We introduce the bandwidth scaling exponent $\psi$ to relate the growth of bandwidth to the number of nodes. Note that $\psi<1$ suggests bandwidth per node decreases as the number of nodes increases, while $\psi>1$ represents asymptotically infinite bandwidth per node.
\begin{table}[!b]
 \centering
 \caption{Scaling Exponents of Network Parameters}
 \label{tab:exponents}
 \begin{tabular}{cc|l}
  Exponent & Range & Parameter (vs. No. of nodes $n$)\\ \hline
  $\psi$ & $[0,\infty)$ & Bandwidth $W=W_0n^\psi$\\
  $\nu$ & $[0,1]$ & Area $A=A_0n^\nu$\\
  $\beta$ & $[0,1]$ & No. of BSs $m=m_0n^\beta$\\
  $\gamma$ & $[0,1-\beta]$ & No. of BS antennas $\ell=\ell_0 n^\gamma$\\
 \end{tabular}
\end{table}

The feasible throughput
definition, adapted from \cite{kumar2003capacity} to the infrastructure case, is given below.
\begin{definition}
 A throughput of $R(n)$ bits per second per node is \textit{feasible} in the
 downlink for a given spatio-temporal scheduling protocol
 in an infrastructure network if all nodes can receive from the BS
 at least $R(n)$ bits per second.
\end{definition}

\begin{figure}[!t]
 \centering
 \includegraphics[width=0.5\columnwidth]{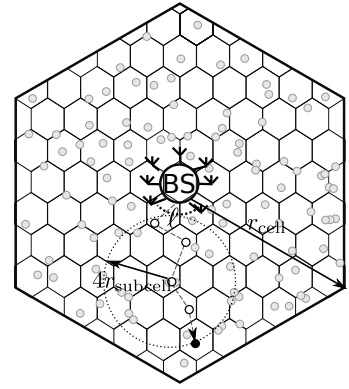}
 \caption{Routing in IMH. Only one cell is shown.}
 \label{fig:modelIMH}
\end{figure}

%
%
In this paper, we do not characterize the capacity scaling of the network, that is the scaling of supremum feasible rate. We focus on illustrating the impact of bandwidth in scaling of two classic protocols. We leave open for future work the discussion of upper bounds, the adaptation of hierarchical cooperation \cite{ozgur2009information} to the cellular case, and the comparison with ad-hoc networks \cite{journals/tit/ShinJDVCLT11}. In particular, we characterize the feasible rate of the following two protocols:
\begin{itemize}
 \item In the \textit{Infrastructure Single Hop} (ISH) protocol, BSs directly transmit to each destination node. Signals that propagate between different cells are treated as interference. There are $\frac{n}{m}$ nodes uniformly distributed within each cell. The $\ell$ BS antennas are used for multi-user MIMO (MU-MIMO): the BS can transmit $\ell$ spatially separated streams at the same time. The feedback channel for channel state information that would be needed to implement the MU-MIMO system is not explicitly taken into account. In each stream $\frac{n}{m\ell}$ nodes are allocated in orthogonal bandwidths $W_{\mathrm{u}}=W\ell\frac{m}{n}$ to each user node $u$. This is always possible because $\gamma<1-\beta$, so there are always more nodes than antennas if $n$ is sufficiently large. Also, the rank of the multi-user channel matrix is at least $\ell$ when nodes are separated at least a quarter of a wavelength and
far-field assumptions hold. The BS transmits with node power allocation $P_{\mathrm{u}}=P_{\mathrm{BS}}\frac{m}{n}$.

 \item In the \textit{Infrastructure Multi Hop} (IMH) protocol, each cell is subdivided regularly into smaller regions of area $A_{\mathrm{r}}$ called \textit{routing sub-cells}, and information is forwarded from the BS via multi-hop communication using a node in each routing sub-cell as relay as shown in Fig \ref{fig:modelIMH}. For multi-hopping, the routing cells must have at least one node with high probability which results in $A_{\mathrm{r}}>\frac{A}{m}\frac{2\log(\frac{n}{m})}{\frac{n}{m}}$ \cite{journals/tit/ShinJDVCLT11}. The BS uses MU-MIMO to start up to $\ell$ routing paths per transmission opportunity at the same time. For the remaining sub-cells, one node forwards data of a single path. Each hop covers distance $d$ bounded by sub-cell radius ($r_{\mathrm{subcell}}$), $d\leq 4r_{\mathrm{subcell}}\propto\sqrt{A_{\mathrm{r}}}$. Sub-cells alternate in becoming active using a non-scaling (i.e. constant) time or frequency division scheduling to avoid collisions and satisfy the half-
duplex constraint.
\end{itemize}

 \begin{remark}
 \label{rem:rem1}
 The use of orthogonal bandwidth allocation in each stream of ISH with MU-MIMO is sufficient to obtain the best feasible rate scaling for ISH. More advanced strategies have higher capacities but the same scaling. This can be shown by modifying the results in \cite{journals/twc/LozanoP12} for the MIMO broadcast channel and arguing that the inner and outer bounds for the capacity region have the same scaling. The full proof is omitted 	due to space limitations.
 \end{remark}


For both protocols we assume that nodes treat out-of- (sub-) cell interference as additional Gaussian noise, and allocations are uniform in frequency so that the links will experience the average interference  Power Spectral Density (PSD). Therefore each receiver experiences an equivalent AWGN with PSD given by
\begin{equation}
\label{eq:defNI}
    N_{\mathrm{I}}=\frac{\sum_{i\in\mathcal{I}} P_{\mathrm{I}}}{W\ell_{\mathrm{t}}}+N_0,
\end{equation}
where $N_0$ is the thermal noise PSD, $W$ is the total bandwidth, $\mathcal{I}$ is the interferer set which depends on the protocol, and $\ell_{\mathrm{t}}$ is the number of transmit-antennas. We use $\ell_{\mathrm{t}}=\ell$ in BS transmissions and $\ell_{\mathrm{t}}=1$ when relays transmit in IMH because the BS divides its power between $\ell$ streams but the amount of interference power captured is not divided. When ${\displaystyle \lim_{n\rightarrow\infty}}\frac{\sum_{i\in\mathcal{I}} P_{\mathrm{I}}}{W\ell_{\mathrm{t}}}=\infty$, $N_I$ is asymptotically interference-limited; when the same limit converges to $0$, it is asymptotically noise-limited and overspreading may occur.

Due to the fact that the protocols separate $\ell$ streams using MU-MIMO (only first hop for IMH), but otherwise use orthogonal bandwidth allocations with fixed power allocations and interference is modeled as AWGN, each transmission in the network can be separately analyzed using the point-to-point results in \cite{journals/twc/LozanoP12}. The following lemma, adapted from \cite{journals/twc/LozanoP12}, describes the feasible rate scaling and critical bandwidth of each transmission in the network as a function only of its allocated bandwidth and power.

\begin{lemma}
\label{lem:Ru}
The orthogonal transmission on the link serving node $u$ achieves scaling
\begin{equation} \label{eq:Rumin}
    R_{\mathrm{u}} =\begin{cases}\Theta\left( W_{\mathrm{u}}\right)&
    W_{\mathrm{u}}<W_{\mathrm{u}}^*\\
    \Theta\left(\frac{P_{r_{\mathrm{u}}}}{N_{\mathrm{I}}} \right)&
    W_{\mathrm{u}}\geq W_{\mathrm{u}}^*\\
    \end{cases}
\end{equation}
where $W_{\mathrm{u}}$ is the bandwidth allocated to the link $u$, that scales as $W_{\mathrm{u}}=W\frac{m\ell}{n}$ for each user in ISH and as $W_{\mathrm{u}}=W$ in each hop of IMH. Also, $P_{r_{\mathrm{u}}}$ is the  average received  power for that link and $W_{\mathrm{u}}^*\propto\Theta\left(\frac{P_{r_{\mathrm{u}}}}{N_I}\right)$ is the critical bandwidth.
\end{lemma}

\section{Results}
\label{sec:DLresults}

This section describes the feasible rate scaling laws for downlink in both protocols. The proofs are relegated to Section \ref{sec:proofs}. It is possible to obtain equivalent theorems for uplink with minor modifications.

\begin{theorem}
 \label{the:ISH}
 Downlink ISH feasible rate per node scales as
 \begin{equation}
  R_{\mathrm{ISH}}(n)\sim
    \Theta\left(n^{\beta-1+\min\left(\psi+\gamma,(\beta-\nu)\frac{\alpha}{2}\right)}\right)
 \end{equation}
\end{theorem}

In ISH, when $\psi+\gamma<(\beta-\nu)\frac{\alpha}{2}$, effective noise PSD ($N_{\mathrm{I}}$ in \eqref{eq:defNI}) becomes asymptotically dominated by interference. Feasible rate of the network is limited by degrees of freedom, determined by number of BSs, transmit antennas and bandwidth. Conversely, when $\psi+\gamma>(\beta-\nu)\frac{\alpha}{2}$, $N_{\mathrm{I}}$ is asymptotically dominated by noise and rate is power-limited with receive power determined by path-loss, node-BS distances and BS density, but not inter-node distance, bandwidth or number of transmit antennas.


\begin{theorem}
 \label{the:IMH}
 Downlink IMH feasible rate per node scales as
 \begin{equation}
  R_{\mathrm{IMH}}(n)\sim
    \Theta\left(n^{\beta-1+\min\left(\psi+\gamma,(1-\nu)\frac{\alpha}{2}\right)}\right)
 \end{equation}
\end{theorem}

In IMH, when $\psi+\gamma<\frac{\alpha}{2}(1-\nu)$, effective noise PSD ($N_{\mathrm{I}}$ in \eqref{eq:defNI}) becomes asymptotically dominated by interference in BS-node links within the BS routing subcell. Feasible rate of the network is limited by degrees of freedom, determined by number of BS, transmit antennas and bandwidth. Conversely, when $\psi+\gamma>(1-\nu)\frac{\alpha}{2}$, $N_{\mathrm{I}}$ is dominated by noise and feasible rate is power-limited with receive power determined by path-loss, inter-node distances and BS density, but not BS-node distance, bandwidth or number of transmit antennas.


The main difference between ISH and IMH is the replacement of BS-node distance with inter-node distances in the power-limited regime. This results in a power gain for IMH that increases the critical bandwidth, postponing the switch from the degrees-of-freedom-limited to the power-limited regime in the multi-hop protocol.
Note that IMH routes start by the BS broadcasting to the nodes in its subcell, and continue by inter-node point-to-point links. In both types of communication power is spread over bandwidth and distances are on the same scale, but in the first one power is also spread between antennas and multiple signals for different receivers, becoming the bottleneck for feasible rate scaling.

\begin{figure}[!ht]
 \centering
 \includegraphics[width=\columnwidth]{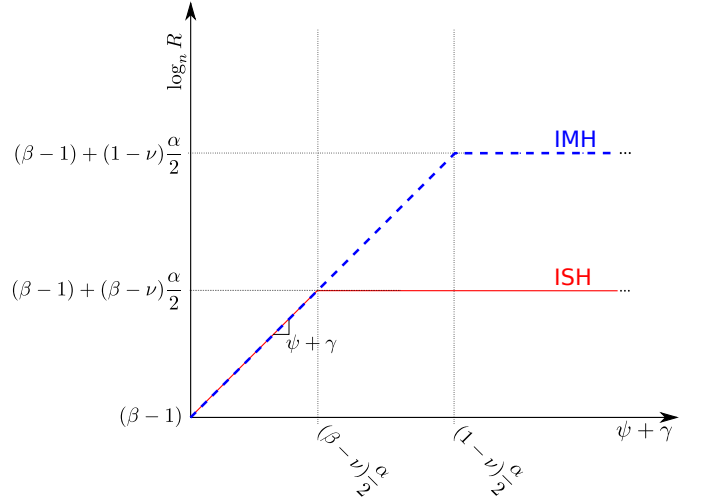}
 \caption{Scaling exponents for downlink ISH and IMH.}
 \label{fig:epsilonDL}
\end{figure}

Figure \ref{fig:epsilonDL} compares the feasible rate scaling for the two protocols. The horizontal axis is the sum of the exponents of bandwidth and number of transmit antennas, $\psi+\gamma$, which represents the scaling of BS transmitted signals degrees of freedom.
The vertical axis represents the exponent of feasible node rate $\log(R(n))$. Infrastructure per node is finite $\beta+\gamma\leq1$. In the degenerate case $\beta=1$, $\gamma=0$, each node can have a non-scaling dedicated channel to a BS, scaling is trivially linear and the protocol is irrelevant. For $\beta<1$ (i.e. increasing
number of users per BS),
the feasible rate scalings in IMH outperforms that of ISH for all bandwidth exponents $\psi+\gamma$.  Note that
\begin{itemize}
\item For $\psi+\gamma<\frac{\alpha}{2}(\beta-\nu)$, IMH and ISH are both degrees-of-freedom-limited. The rates of the two protocols do not differ in terms of scaling.
\item For $\psi+\gamma>\frac{\alpha}{2}(\beta-\nu)$, IMH strictly surpases ISH. In the range $\frac{\alpha}{2}(\beta-\nu)<\psi+\gamma<\frac{\alpha}{2}(1-\nu)$ IMH remains degrees-of-freedom-limited although ISH is power-limited. The rate of IMH still grows with the bandwidth exponent, but this is not true for ISH.
\item For $\psi+\gamma>\frac{\alpha}{2}(1-\nu)$, both IMH and ISH are power limited. The rate of IMH ceases to grow with the bandwidth exponent but its exponent is larger than that of ISH since IMH obtains a power gain due to shorter transmission range.
\end{itemize}

\section{Proofs}
\label{sec:proofs}
\subsection{Proof of Theorem 1}
\label{sec:proofsISHDL}

The proof relies on the following two ideas
\begin{enumerate}
 \item The exponent of $N_{\mathrm{I}}$ for ISH is
 \begin{equation}
 \label{eq:NIDL}
  N_{\mathrm{I,ISH}}\sim\Theta\left(n^{\max\left(\frac{\alpha}{2}(\beta-\nu)-\psi-\gamma,0\right)}\right)
 \end{equation}
 \item In ISH asymptotically all links become degrees-of-freedom limited when $N_{\mathrm{I}}$ is interference limited, and power limited when it is noise limited. In other words, when $N_{\mathrm{I}}$ is dominated by interference (noise) the probability of overspreading tends to zero (one) as $n\rightarrow\infty$.
\end{enumerate}

To prove the first point, we consider for each node $u$ the interferer set $\mathcal{I}_{\mathrm{ISH}}$ containing all BSs except the one that serves $u$. We get
\begin{equation}
\label{eq:NIISHDL}
N_{\mathrm{I,ISH}}=\frac{1}{W\ell}\sum_{i\in\mathcal{I}_{\mathrm{ISH}}}r_{i,u}^{-\alpha}P_{\mathrm{BS}}+N_0.
\end{equation}
where $r_{i,u}$ is the distance from user $u$ to BS $i$,. Note that we can lower bound $r_{i,u}$ by the distance between $i$ and the border of the cell. In the hexagonal tessellation of the plane there are $6k$ cells that form a ring at exactly $2k-1$ cell radii from node $u$. The network is finite and a maximum $k$ exists, but we can get rid of border effects by extending the sum of these interfering rings by having $k\rightarrow\infty$.
\begin{equation}
\begin{split}
  \sum_{i\in\mathcal{I}_{\mathrm{ISH}}}r_{i,u}^{-\alpha}
    &\leq r_{\mathrm{cell}}^{-\alpha}\sum_{k=1}^{\infty}(6k)(2k)^{-\alpha}\\
    &\leq 6(2r_{\mathrm{cell}})^{-\alpha}\zeta(\alpha-1)\\
\end{split}
\end{equation}
where $\zeta(\alpha-1)$ is the Riemann Zeta function evaluated in $\alpha-1$, which is just some constant for any fixed $\alpha>2$.
This shows that interference power scales as $n^{\frac{\alpha}{2}(\beta-\nu)}$, while noise PSD is constant, so \eqref{eq:NIISHDL} scales as \eqref{eq:NIDL}.

To prove the second point we use the definition of critical bandwidth $W^*$ in \cite{journals/twc/LozanoP12}. We express the critical bandwidth for each user $u$ as a function of its distance from the BS $r_{\mathrm{u}}$, received power of the user $P_{\mathrm{u}}=P_{\mathrm{BS}}\frac{m}{n}$, and relative fraction of bandwidth allocated to each user $\frac{m}{n}\ell$.
\begin{equation}
 W^*(r_{\mathrm{u}})\propto \frac{P_{\mathrm{BS}}}{\ell N_I}(r_{\mathrm{u}})^{-\alpha}
 \end{equation}
Instead of comparing the bandwidth limitation $W^*(r_{\mathrm{u}})$ and the actual bandwidth $W$ one user at a time, we compute the \textit{critical distance} from the BS, $r^*$, that marks the border between the region where users do not see overspread transmissions and where they observe overspread transmissions with the allocated power and bandwidth.
The critical distance of ISH scales as
\begin{equation}
 r^*_{\mathrm{ISH}}\sim\Theta(n^{\frac{(-\psi-\gamma)}{\alpha}}).
\end{equation}

We divide the cell in two areas: an inner circle containing nodes at less than the critical distance and an outer ring containing the nodes at more than the critical distance.
%
%
The fraction of nodes inside the circle
\begin{equation}
\label{eq:FIU}
 f_{\mathrm{ISH}}=\min\left(\frac{2\pi (r^*_{\mathrm{ISH}})^2}{A_0n^{\nu-\beta}},1\right)\propto n^{\frac{-2(\gamma+\psi)}{\alpha}+(\beta-\nu)},
\end{equation}
converges to one when $\frac{-2(\gamma+\psi)}{\alpha}+(\beta-\nu)>0$ (interference-limited case), and to zero otherwise (noise-limited case).

The rates feasible in each regime are given by Lemma \ref{lem:Ru}: $\Theta(W_{\mathrm{u}})=\frac{m\ell}{n}W$ in the degrees-of-freedom-limited case and $\Theta(\frac{P_{r_{\mathrm{u}}}}{N_{\mathrm{I}}})\propto n^{\beta-1+\frac{\alpha}{2}(\beta-\nu)}$ in the power-limited case. Putting everything together we have Theorem \ref{the:ISH}.

\subsection{Sketch of Proof of Theorem 2}
\label{sec:proofsIMHDL}

In the first hop of each route, the BS-node feasible rate can be analyzed similarly to ISH replacing $\beta=1$, $W_\mathrm{u}=W$ and $P_{\mathrm{u}}=\frac{P_{BS}}{\ell}$. Further hops in each route are point-to-point links with single antenna, hence following a similar analysis results in scaling as in ISH replacing $\gamma=0$, $\beta=1$. By combining the analysis of different hops we can distinguish three regimes:
\begin{itemize}
 \item $\psi+\gamma<\frac{\alpha}{2}(1-\nu)$, all links are degrees-of-freedom-limited.
 \item $\psi<\frac{\alpha}{2}(1-\nu)<\psi+\gamma$, first-hop links are power-limited, the following are interference limited. The fraction of nodes that are not overspread in the first hop converges to zero and overspreading affects the allocation of $\ell W$ antennas and bandwidth resources at the BS, but not usage of bandwidth $W$ when nodes transmit.
 \item $\psi>\frac{\alpha}{2}(1-\nu)$, all links are power-limited. The fraction of nodes that are not overspread nodes in second and following hops converges to zero and overspreading affects all uses of bandwidth.
\end{itemize}

Using Lemma \ref{lem:Ru}, feasible rates in the first hop are $\Theta(W_{\mathrm{u}})=\frac{m\ell}{n}W$ in the first regime and $\Theta(\frac{P_{r_{\mathrm{u}}}}{N_{\mathrm{I}}})\propto n^{\beta-1+\frac{\alpha}{2}(1-\nu)}$ in the other two.

In the second and further hops, feasible rates are $\Theta(W_{\mathrm{u}})=W$ in the two first regimes, and $\Theta(\frac{P_{r_{\mathrm{u}}}}{N_{\mathrm{I}}})\propto n^{\frac{\alpha}{2}(1-\nu)}$ in the last regime. However, these rates are obtained per route and not per node, so they must normalized multiplying by $\ell$ simultaneous routes and dividing by $\frac{n}{m}$ nodes per cell.

Comparing the rates in all regimes shows that the bottleneck is always the first hop. Combining, we obtain Theorem \ref{the:IMH}.

\section{Conclusions and Future Work}
\label{sec:conclusion}

We have studied scaling laws of infrastructure cellular networks subject to overspreading effects with simultaneous scaling of nodes and bandwidth. Traditional cellular architectures with only direct transmissions between base stations and mobile
nodes become power-limited in wideband regimes, unable to fully utilize the bandwidth.
In contrast, multi-hop systems may be able to
better utilize large bandwidth
or number of antenna degrees of freedom, provided the number of nodes per base station
is increasing.
Although multi-hop systems have demonstrated limited value in current
bandwidth-limited cellular systems, our finding suggest that
multi-hop transmissions deserve reconsideration for
emerging mmW 5G networks where wide bandwidths and large number of antennas
are available.

Our ongoing research investigates the scaling of capacity upper-bounds and feasible rate of more advanced protocols, the extension of the wideband scaling model to ad-hoc networks, and the integration in new ad-hoc versus infrastructure comparisons taking into account wideband effects.


\bibliographystyle{IEEEtran}
\bibliography{WidebandScaling}

\end{document}